\documentclass[10pt,twocolumn,aps,prl,floatfix,superscriptaddress]{revtex4-2}
\usepackage{graphicx}
\usepackage{color}
\usepackage{amsmath}
\usepackage{amssymb}
\usepackage{comment}
\usepackage{hyperref}

\begin{document}
\title{Identification of a telecom wavelength single photon emitter in silicon}
\author{P\'eter Udvarhelyi}
\affiliation{Wigner Research Centre for Physics, P.O.\ Box 49, H-1525 Budapest, Hungary}
\author{B\'alint Somogyi}
\affiliation{Wigner Research Centre for Physics, P.O.\ Box 49, H-1525 Budapest, Hungary}
\author{Gerg\H{o} Thiering}
\affiliation{Wigner Research Centre for Physics, P.O.\ Box 49, H-1525 Budapest, Hungary}
\author{Adam Gali}
\affiliation{Wigner Research Centre for Physics, P.O.\ Box 49, H-1525 Budapest, Hungary}
\affiliation{Budapest University of Technology and Economics,  Budafoki \'ut 8, H-1111 Budapest, Hungary}

\begin{abstract}
We identify the exact microscopic structure of the G photoluminescence center in silicon by first principles calculations with including a self-consistent many-body perturbation method, which is a telecommunication wavelength single photon source. The defect constitutes of $\text{C}_\text{s}\text{C}_\text{i}$ carbon impurities in its $\text{C}_\text{s}-\text{Si}_\text{i}-\text{C}_\text{s}$ configuration in the neutral charge state, where $s$ and $i$ stand for the respective substitutional and interstitial positions in the Si lattice. We reveal that the observed fine structure of its optical signals originates from the athermal rotational reorientation of the defect. We attribute the monoclinic symmetry reported in optically detected magnetic resonance measurements to the reduced tunneling rate at very low temperatures. We discuss the thermally activated motional averaging of the defect properties and the nature of the qubit state.
\end{abstract}

\maketitle

Emerging material platforms realizing single photon emitters and spin-photon interfaces  are essential for quantum telecommunication applications~\cite{Zhang2020}. Interfacing the local spin qubits of a point defect to photon qubits capable for long distance coherent transmission in optical fibers provides the basis of creating quantum internet. Besides the already successful quantum emitters in diamond and silicon carbide, promising single photon emitter defects have been created and measured in silicon recently~\cite{Hollenbach2020, Redjem2020, Durand2020}, some of them associated with the so-called G photoluminescence center~\cite{Bean1970} which emits in the telecom O-band. As silicon is the most mature material in terms of unprecedented fabrication capabilities for electronics and photonics structures, the recently discovered single photon source G-center with telecom wavelength may turn silicon to such a quantum-coherent material which unifies the electronics, photonics, and quantum optics components into a single, completely integrable platform. To this end, the magneto-optical properties of G-center should be understood in great detail to unravel quantum optics protocols.

\begin{figure}
    \centering
    \includegraphics[scale=0.5]{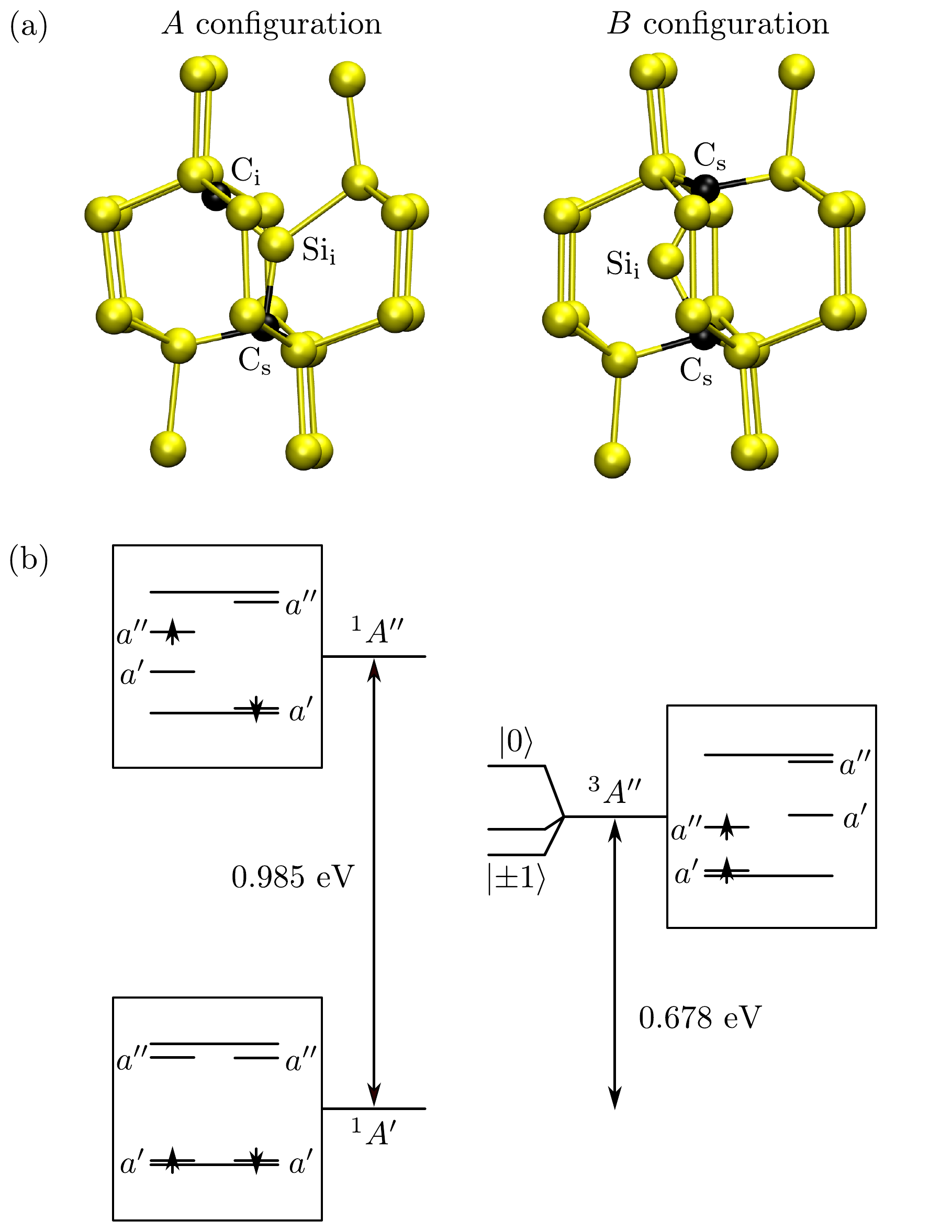}
    \caption{\label{fig:geom}%
    (a) The structure of the $\text{C}_\text{s}\text{C}_\text{i}$ bistable defect in silicon. The $B$ configuration is suggested for the G-center by experiments, where the atoms are rearranged to form the $\text{C}_\text{s}-\text{Si}_\text{i}-\text{C}_\text{s}$ structure. (b) Visualization of the many-electron and single electron levels of the $\text{C}_\text{s}\text{C}_\text{i}$ defect in $B$ configuration in silicon. The band gap is represented by horizontal lines corresponding to band edge states, the spin polarized defect levels are separated to two spin channels. Excitation energies are calculated with HSE06+U $\Delta$SCF method and using exchange correction (see main text).}
\end{figure}

The G-center has been extensively investigated by various experiments over half a century. The zero-phonon-line (ZPL) of G-center appears in carbon-rich silicon at 0.97~eV~\cite{Bean1970,Thonke1981} which is associated with a damage center consisting of two carbon impurity atoms. Uniaxial stress measurements of the G-center showed monoclinic ($\text{C}_\text{1h}$) symmetry~\cite{Foy1981}.
Optically detected magnetic resonance (ODMR) in its metastable triplet state was also observed~\cite{Lee1982}.  Two configurations of the defect showing monoclinic symmetry, labeled $A$ and $B$, are proposed by deep-level transient capacitance spectroscopy (DLTS) and electron paramagnetic resonance (EPR) measurements~\cite{Jellison1982, Song1990}. The defect shows bistability, its $\pm 1$ charge state is stable in $A$ configuration and its neutral charge state is stable in $B$ configuration. The suggested structure of the former consists of a carbon-silicon split-interstitial pair and a neighboring substitutional carbon atom ($\text{C}_\text{i}\text{Si}_\text{i}-\text{C}_\text{s}$), whereas the latter can be described by two substitutional carbon atoms and a silicon interstitial between them ($\text{C}_\text{s}-\text{Si}_\text{i}-\text{C}_\text{s}$), which is distorted from the $\left<111\right>$ bond axis to $\text{C}_\text{1h}$  symmetry~\cite{ODonnell1983}  [see Fig.~\ref{fig:geom} (a)]. The G photoluminescence line arises only from the proposed $B$ configuration in the neutral charge state. ODMR studies at T=1.7~K showed a monoclinic symmetry of the defect~\cite{Lee1982} with motional averaging at T=30~K, whereas trigonal symmetry was observed at T=5~K by another ODMR study~\cite{Yan1986}, corroborated by an EPR study recorded at T=6~K~\cite{Vlasenko1995}. Recently, fine structure in the ZPL of absorption spectrum has been observed with 1-2-2-1 degeneracy and energy separations of $\delta:2\delta:\delta$ ratio with $\delta=2.5~\mathrm{\mu eV}$,  in highly $^{28}\text{Si}$ enriched sample at T=1.4~K~\cite{Chartrand2018}. Isotope shifts in the fine spectrum have been also reported in this study. The spectral lines are completely broadened at T=20.0~K with an activation energy of 12.4~meV~\cite{Chartrand2018}. {\color{black}No physical model from first principles theory~\cite{Mattoni2002, Liu2002, Zirkelbach2011, Docaj2012, Wang2014SR, Wang2014, Timerkaeva2018} was provided for these observations in relation to the previous findings (see Supplemental Material in Ref.~\footnote{See Supplemental Material at}), thus no clear consensus has been reached about the exact origin of the defect which is the first inevitable step towards understanding their properties.}

In this Letter, we unambiguously identify the microscopic structure of the defect {\color{black} based on the proposed models} by calculating all the spectral fingerprints from experiments, i.e., the optical transitions energy between its singlet states, the fine structure of the optical signal and the zero-field-splitting and hyperfine interaction in its metastable triplet state as well as fine structure and isotope shift in the optical signal. After identification of the G-center, we model the athermal reorientation of the defect and explain the thermal averaging observed in optically detected magnetic resonance measurements. Our calculations reveal the exotic quantum properties of the defect, e.g., the spin-rotation coupling, and provide guidance for future experiments to control the qubit state.

The structural model of the defect is created in a 512-atom silicon supercell and relaxed with density functional theory (DFT) calculations, using Heyd-Scuseria-Ernzerhof (HSE06) functional~\cite{HSE06} and a single $\Gamma$-point sampling of the Brillouin zone, as implemented in the VASP plane wave based code~\cite{VASP1,VASP2,VASP3,VASP4}. Excited states are calculated with $\Delta$SCF method~\cite{Gali2009}. Hyperfine and ZFS parameters are calculated with the VASP implementation of Martijn Marsman~\cite{Szasz2013, Bodrog2013}. {\color{black}We show that the accurate description of spin-spin interactions in the defect requires a local correction with structure optimization~\cite{Cococcioni2005, Janotti2006, Kovacik2009, Meredig2010, Ivady2013, Ivady2014} by applying a Hubbard U onsite potential in the Dudarev approach~\cite{Dudarev1998} on the p-orbital of the frustrated Si self-interstitial atom (see the details on the HSE06+U calculations in Sec.~II in~\onlinecite{Note1}). Detailed description of the methods, the formation and stability of G-center will be given in a forthcoming work.}

{\it Electronic structure and optical excitation}.---
Next, we discuss the results of HSE06+U calculations on the electronic structure of the $\text{C}_\text{s}\text{C}_\text{i}$ defect. In the following, we discuss only the most stable $B$ configuration in the neutral charge state with the widest region of stability (see the results for charged defects with charge correction~\cite{Freysoldt2009, Wickramaratne2018} in Sec.~III in \onlinecite{Note1}). In its ground state, the defect introduces a fully occupied level resonant with the valence band edge ($a^{\prime}$) and an empty level ($a^{\prime\prime}$) in the band gap [see Fig.~\ref{fig:geom} (b)]. These orbitals are strongly localized on the self-interstitial silicon atom but in a heavily frustrated sp-like configuration which is highly atypical bonding configuration for Si atoms. The ground state total electron configuration is $^{1}A^{\prime}$. Promoting an electron from the $a^{\prime}$ to the $a^{\prime\prime}$ level lifts the hole level into the band gap, creating $^{1}A^{\prime\prime}$ and $^{3}A^{\prime\prime}$ excited states. The $\text{C}_\text{1h}$ point group of the defect defines the spin quantization axis perpendicular to the mirror plane. Parallel and perpendicular directions are referenced to this quantization axis. Here we note that the triplet excited state of the defect is stable in $\text{C}_1$ symmetry, connected to the $\text{C}_\text{1h}$ configuration by dynamical reorientation of the defect (discussed below). However, this symmetry breaking has a minor effect on the extent of the defect levels, thus we label them according to the $\text{C}_\text{1h}$ symmetry counterparts. Allowed optical transition between the two singlet states with parallel transition dipole moment $d_{\parallel}$ originates the G line. The DFT calculated total energy difference in the $^{1}A^{\prime}\leftrightarrow {^{1}}A^{\prime\prime}$ and $^{1}A^{\prime}\leftrightarrow {^{3}}A^{\prime\prime}$ transitions are 0.985~eV and 0.678~eV, respectively. The former contains a correction to accurately describe the open-shell singlet excited state~\cite{Ziegler1977}, resulting in a good agreement with the experimental ZPL energy.

{\it Spin properties}.---
ODMR was demonstrated in the metastable triplet state of the G-center~\cite{Lee1982, ODonnell1983}.
The main contribution to the D-tensor in our DFT calculation originates from the localized defect orbital on the central silicon atom. We compare the DFT results with the applied onsite correction and the experimental spin coupling parameters in Table~\ref{tab:hyperfine}. The calculated parameters of the $\text{C}_\text{s}\text{C}_\text{i}$ defect in the $B$ configuration are in reasonable agreement with experimental findings in the G-center. For the comparison of the results obtained with and without the applied $U$ correction, see Sec.~IV in \onlinecite{Note1}. ODMR measurements reported thermally activated reorientation in the ${^{3}}A^{\prime\prime}$ state~\cite{Lee1982, ODonnell1983, Yan1986}. We also provide the calculated axial $D_\text{avg}$ motionally averaged parameter by averaging the D-tensor for the equivalent defect positions.

\begin{table}
    \centering
    \caption{Comparison of the hyperfine parameters of the defect $^{29}\text{Si}$ atom and the zero field splitting parameters in the G center. Experimental (exp.) monoclinic $D$ eigenvalues and the axial $D_\text{avg}$ are taken from Ref.~\citenum{ODonnell1983} and Ref.~\citenum{Yan1986}, respectively. The axial $D_\text{avg}$ motionally averaged parameter is calculated by averaging the D-tensor for the equivalent defect positions. DFT values are calculated with HSE06 functional with a Hubbard U correction for the defect orbitals. The hyperfine values contain the core polarisation contribution.}
    \begin{ruledtabular}
    \begin{tabular}{ccc}
         parameter & exp. (MHz) & HSE06+U (MHz) \\\hline
         $A_{zz}$ & 339   & -347 \\
         $A_{yy}$ & 312 &  -324\\
         $A_{xx}$ & 273 & -267\\
         $D_{zz}$ & $\pm$941 & -1218 \\
         $D_{yy}$ &  $\pm$800 &  911\\
         $D_{xx}$ &  $\pm$142 &  307\\
         $D_\text{avg}$ & 1210 & 1365
    \end{tabular}
    \end{ruledtabular}
    \label{tab:hyperfine}
\end{table}

 {\it Rotational motion of the defect}.---
 In the $\text{C}_\text{s}\text{C}_\text{i}$ defect, the central silicon interstitial of the defect is twofold coordinated, thus a thermal averaging to $\text{C}_\text{3v}$ symmetry was observed in the triplet state at elevated temperatures~\cite{Foy1981, Lee1982, ODonnell1983}. This process is plausible in the ground and excited singlet states as well. As the central interstitial silicon atom of the defect is strongly bound to only the two carbon neighbors its motion in the plane perpendicular to [111] direction is possible with relatively low barrier energy $V_0$ [see Supplemental Material~\cite{Note1} Fig.~4 (a) and (b)]. The back-bonds of the carbon neighbor atoms designate two sets of threefold degenerate minima of the potential energy. So the reorientation takes place in two planes with threefold symmetry, however the separation of the planes is very small, only 0.062~\AA. The most important features of the defect reorientation can be modeled in a higher $\text{D}_{3\text{d}}$ symmetry and the potential energy function can be approximated with a sixfold symmetric periodic well. In order to parametrize the function, we perform nudged elastic band (NEB) calculation (see Ref.~\cite{Note1} for technical details).

\begin{figure}
    \centering
    \includegraphics[scale=0.5]{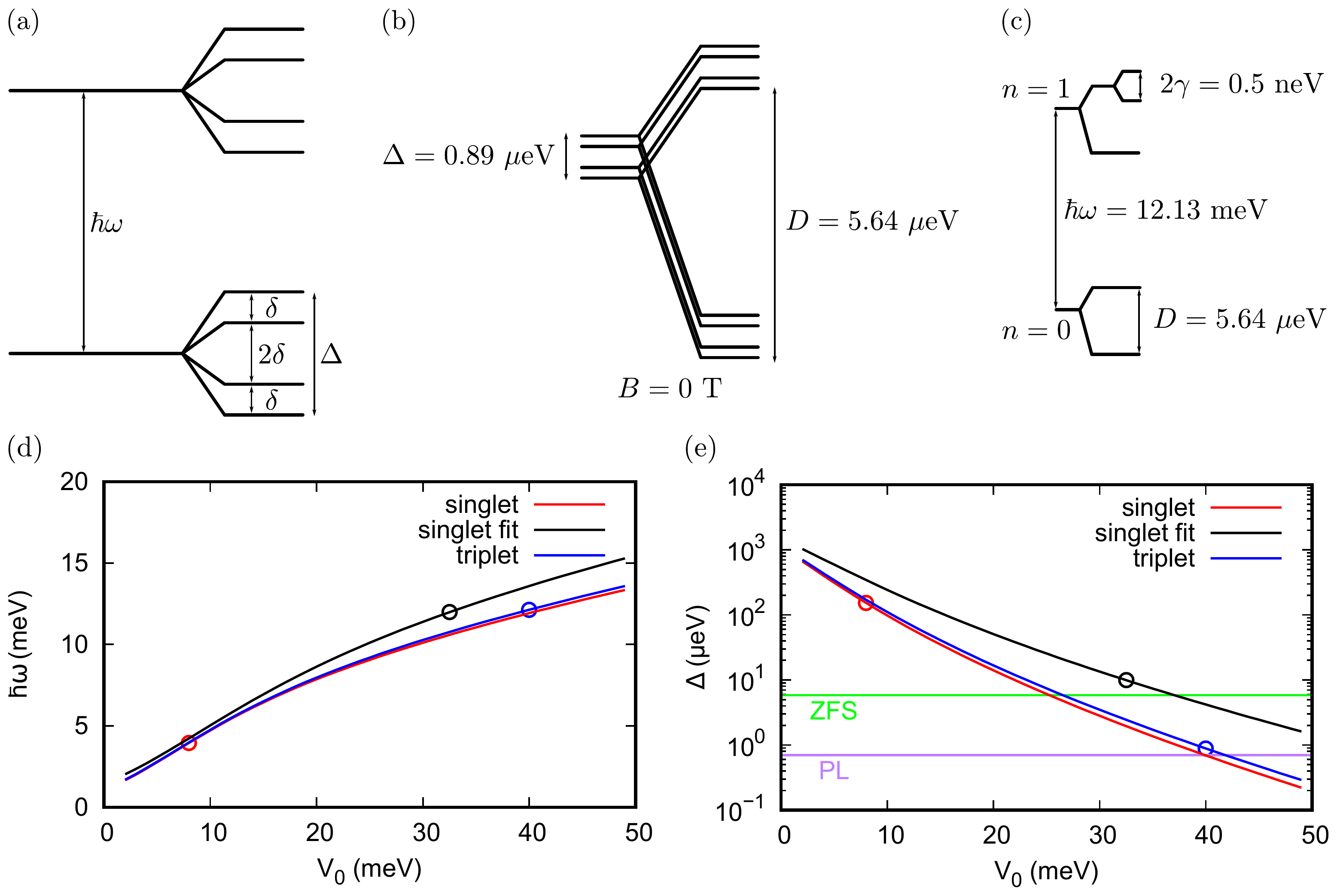}
    \caption{\label{fig:tunneling}%
    (a) Energy level diagram in the periodic cosine potential with six minima, corresponding to the rotational reorientation in the triplet state of the $\text{C}_\text{s}\text{C}_\text{i}$ defect. The diagram is combined with the zero-field splitting (ZFS), shown only in the excited rotational state for clarity. The characteristic energies are the oscillator quantum $\hbar\omega$ the tunneling splitting energy $\Delta$ and the motionally averaged ZFS parameter $D$.
    (b) Oscillator energy and (c) tunneling-splitting energy in the function of the reorientation barrier energy. The corresponding values in the singlet and triplet excited states are marked with circles. Characteristic frequencies of the photoluminescence (PL)~\cite{Beaufils2018} and zero-field splitting (ZFS)~\cite{Lee1982} are also marked. The singlet curve is fitted to reproduce the experimental results in Ref.~\onlinecite{Chartrand2018}.}
\end{figure}

The HSE06+U NEB calculations for the ground state resulted in $V_0=89~\mathrm{meV}$ barrier energy for rotation and long tunneling path at $Q=31.97~\sqrt{\mathrm{u}}\text{\AA}$. As a consequence of the large $V_0$, the splitting between the levels are negligible and the spectrum is sixfold quasi-degenerate. We determined these parameters for the triplet excited state too for which the adiabatic potential energy surface (APES) can be reliably mapped by $\Delta$SCF method. The length of the total path is $Q=25.7~\sqrt{\mathrm{u}}\text{\AA}$ and the barrier energy for rotation is $V_0=40~\mathrm{meV}$. Here we note that the minimum potential energy positions along the circular tunneling path is rotated by $30^{\circ}$ compered to the ground and excited state positions. Thus the energy minima and barriers in the triplet excited state belong to $\text{C}_1$ and $\text{C}_{1\text{h}}$ point symmetry, respectively. For finite potential barrier energy, the sixfold degeneracy is partially split into a quartet structure with 1-2-2-1 degeneracy and energy separations of $\delta-2\delta-\delta$ sequence, where $\delta$ corresponds to the tunneling rate through a single barrier and the total tunneling splitting is $\Delta=4\delta$ [see Fig.~\ref{fig:tunneling}(a)]. The same quartet fine structure in the fluorescence spectrum with $\delta=2.5~\mathrm{\mu eV}$ has been recently observed in highly $^{28}\text{Si}$ enriched sample~\cite{Chartrand2018}. The observed splitting can only be attributed to the tunneling splitting of the singlet excited state not to the ground state. The observed activation energy in Ref.~\onlinecite{Chartrand2018} was around $12~\mathrm{meV}$. We find that the value of $\hbar\omega$ in the calculated triplet state well matches this energy gap and we assume a similar value for the singlet excited state because of the similar electronic configuration, thus the activation energy can be explained by phonon excitation rather than electronic excitation, in contrast to the suggestion in Ref.~\onlinecite{Chartrand2018}. For the singlet excited state, we finally fit the $Q$ and $V_0$ parameters to the cited experimental $\hbar\omega$ and $\delta$ data (see the explanation and details in Ref.~\onlinecite{Note1}) which results in $Q=22.5~\sqrt{\mathrm{u}}\text{\AA}$ and $V_0=33$~meV.

With these parameters, we calculate the isotope effect on the rotational reorientation by scaling the length of the tunneling path with the corresponding changes in the atomic masses. The calculated isotope shifts for the central silicon atom are $54~\mathrm{\mu eV}$ and $106~\mathrm{\mu eV}$ for $^{29}\text{Si}$ and $^{30}\text{Si}$, respectively, in excellent agreement with the experimental data ($\sim50$ and $\sim100$~$\mu$eV, respectively, in Ref.~\onlinecite{Chartrand2018}).

{\it Temperature dependent tunneling}.---
At $0~\mathrm{K}$, the dynamics of the system is governed by the tunneling splitting $\Delta$. For the specific barrier energy in the singlet excited state, the optical lifetime in the PL measurement is longer than the characteristic time of tunneling [see Fig.~\ref{fig:tunneling}(c)]. Therefore, the motional averaging results in a high symmetry ($\text{D}_{3\text{d}}$) rotational configuration and the tunnelling splitting can be observed in the PL spectrum~\cite{Chartrand2018}.
The smaller athermal reorientation frequency of $\Gamma_{0}=6\delta/h=0.321~\mathrm{GHz}$ calculated in the triplet state indicate monoclinic symmetry in the ZFS at zero temperature.

On the other hand, the tunneling rates can be enhanced at elevated temperatures assisted by acoustic phonons. The interaction with phonons can be described beyond the Born-Oppenheimer approximation, where the elastic distortions associated with the acoustic phonons perturb the APES~\cite{Nalbach2017}. Treating this electron-phonon interaction as a time dependent perturbation of the athermal tunneling solution introduces temperature dependent direct ($\propto T$) and Raman ($\propto T^5$) contributions to the rotational tunneling rates (see Refs.~\onlinecite{Jahnke2015, Goldman2015} and Sec.~V in \onlinecite{Note1} for further details)
\begin{equation}
    \Gamma(T)=\Gamma_{0}+\Gamma_\text{direct}+\Gamma_\text{Raman}=\frac{6\delta}{h}+\alpha(\delta)T+\beta(\delta)T^5\text{,}
\end{equation}
where $\alpha(\delta)$ and $\beta(\delta)$ functions incorporate the electron-phonon coupling strength, the density of phonon states and constants. The calculation of these functions are beyond the scope of this Letter. 
As the activation energy ${\delta=0.22~\mathrm{\mu eV}}\approx 2.55~\mathrm{mK}\cdot k_\text{B}$  in the triplet state ($k_\text{B}$ is the Boltzmann constant), we estimate two phonon Raman transitions to dominate in the $T> 1~\mathrm{K}$ region.

The ODMR measurements of Lee {\it et al.}~\cite{Lee1982} was performed at $T=1.7~\mathrm{K}$ in a $35~\mathrm{GHz}$ $\text{TE}_{011}$ microwave cavity and the ODMR spectrum was recorded by sweeping the $[011]$ aligned external magnetic field. They also reported preliminary studies using $35~\mathrm{GHz}$ microwave frequency at elevated temperatures ($\sim30~\mathrm{K}$) showing thermal averaging. EPR~\cite{Vlasenko1995} and ODMR\cite{Yan1986} measurements reported trigonal symmetry at $6~\mathrm{K}$ and $5~\mathrm{K}$, respectively, within the same order of interrogation frequency as used above. These results imply that the thermally activated reorientation is significant at very low temperatures. We estimate that the Raman term with a large $\beta(\delta)$ prefactor enhances the rate of reorientation above the interrogation frequency around $5~\mathrm{K}$. This steep thermal activation should be significant at $T=1.7~\mathrm{K}$ as well. Thus we anticipate that the thermal averaged ODMR signal may be observable without applied external magnetic field.

{\it The nature of the singlet excited state and the triplet qubit state}.---
Our calculations revealed that the $2.5~\mathrm{\mu eV}\approx 0.6~\mathrm{GHz}$ splitting in the fine structure of the ZPL energies is associated with the rotational levels of the interstitial Si-atom in the singlet excited state, and the isotope shifts upon substituting the $^{28}$Si to heavier isotopes of the interstitial Si-atom can well explain the shift in of the ZPL lines. The athermal reorientation of the defect also occurs in the triplet qubit state but at slower rate. Assuming that the rotational states and the spin subspace are decoupled, the same rotational levels appear in the fine structure of the ZFS. The combined tunneling-splitting and motionally averaged zero-field-splitting structure at zero external magnetic field is depicted in Fig.~\ref{fig:tunneling}(a). This picture is slightly perturbed by the inclusion of spin-rotational coupling (see Sec.~VI in \onlinecite{Note1}). As our calculations predict a large energy separation of the triplet level from the singlet ground and excited states, the usual ODMR mechanism through spin selective intersystem-crossing via spin-orbit coupling assisted by phonons would not be efficient. Instead, we propose that ionization from above band gap excitation should play a significant role in the spin selective singlet-triplet transition of the defect, presumably, with interaction of other paramagnetic defects.

Our calculations identified the microscopic structure of G-center in silicon. This is the first step in the tight control for the formation of the defect and in-depth characterization of their magneto-optical properties. We could identify the energy position of the metastable triplet level between the singlet levels as well as the spin levels in the triplet manifold that is crucial in the optical control and pumping to the qubit state of the defect. The G-center in silicon exhibits very interesting physics where rotational, orbital, isotope mass with nuclear spin and electron spin degrees of freedom are coupled, also as a function temperature, which can be basically controlled by optical means. Electrical control of emission and spin readout is in reach as the (spin-dependent) optical response was observed by above-band-gap illumination which generates free carriers in silicon. We propose that G-center has a potential as a qubit in silicon but it requires a tight control of free carriers in the crystal and bound exciton states of the defect (e.g., Ref.~\onlinecite{Zhang2020_prl} and see Sec.~VII in \onlinecite{Note1} for further discussion).

We acknowledge that the results of this research have been achieved
using the DECI resource Eagle HPC based in Poland at Poznan with
support from the PRACE aisbl and resources provided by the Hungarian Governmental Information Technology Development Agency. A.~G.\ acknowledges the National Research, Development, and Innovation Office of Hungary grant No.\ KKP129866 of the National Excellence Program of Quantum-coherent materials project, No.\ 2017-1.2.1-NKP-2017-00001 of the National Quantum Technology Program, and the Quantum Information National Laboratory supported by the Ministry of Innovation and Technology of Hungary, as well as the EU Commission for the H2020 Quantum Technology Flagship project ASTERIQS (Grant No.\ 820394).

\nocite{LDA1, LDA2, PBE, Hedin1965, Strinati1988, Faleev2004, scGW, Nicholls_1979, Watkins_1996}

%

\end{document}